\documentclass[
floatfix,
reprint,
superscriptaddress,
 amsmath,amssymb,
 aps,
 pra,
]{revtex4-2}

\usepackage{graphicx}

\usepackage{dcolumn}

\usepackage{bm}

\usepackage{hyperref}

\usepackage{comment}

\usepackage{physics}

\usepackage{float}

\usepackage{numprint}

\usepackage{xifthen}

\usepackage[svgnames]{xcolor}

\usepackage{soul}

\usepackage{epstopdf}

\usepackage{multirow}

\usepackage{titlesec}

\usepackage{multirow}
\usepackage[normalem]{ulem}

\graphicspath{figures}
\graphicspath{{figures/}}

\definecolor{rulecolor}{RGB}{0,71,171}

\definecolor{tableheadcolor}{gray}{0.92}

\colorlet{tmpcolor}{black}

\colorlet{changecolor}{tmpcolor}
\colorlet{CHANGECOLOR}{tmpcolor}

\setstcolor{changecolor}

\hypersetup{
    colorlinks=true,
    linkcolor=blue, 
    citecolor=blue,
    urlcolor=blue
}

\DeclareUnicodeCharacter{3000}{\textcolor{red}{BAD!!}}

\DeclareUnicodeCharacter{2009}{\textcolor{red}{BAD!!}}

\DeclareUnicodeCharacter{03A3}{$\Sigma$}

\DeclareUnicodeCharacter{03A0}{$\Pi$}

\DeclareUnicodeCharacter{03BC}{$\mu$}

\newcommand{\rfig}[1]{Fig.~\ref{#1}}

\newcommand{\rsec}[1]{Section~\ref{#1}}

\newcommand{\req}[1]{Eq.~(\ref{#1})}

\newcommand{\rtab}[1]{Tab.~\ref{#1}}

\newcommand{\mycancel}[1]{{\protect\st{#1}}}
\renewcommand{\mycancel}[1]{}

\newcommand{\remove}[1]{\ignorespaces}

\newcommand{\musec}{\textmu s}

\newcommand{\company}[1]{\ignorespaces}

\newcommand{\artiq}{ARTIQ}

\newcommand{\be}[1]{
\begin{eqnarray}#1
}

\newcommand{\ee}{
\end{eqnarray}
}

\newcommand{\ic}{i\mkern1mu}

\renewcommand{\imath}{\ic}

\begin{document}

\title{Monolithic printed-circuit board RF-trap for 
electrons}

\author{Zijue Luo}%
\affiliation{Department of Physics and Astronomy, University of California, Riverside, California 92521, USA}

\author{Jae Eu}%
\affiliation{Department of Physics and Astronomy, University of California, Riverside, California 92521, USA}

\author{Tianyi Wang}%
\affiliation{Department of Physics and Astronomy, University of California, Riverside, California 92521, USA}

\author{Boerge Hemmerling}%
\email{boergeh@ucr.edu}
\affiliation{Department of Physics and Astronomy, University of California, Riverside, California 92521, USA}

\date{\today}

\begin{abstract}
Qubits encoded in the spin of trapped electrons have been proposed as a promising novel platform for quantum information processing.
While trapping of electrons has been largely carried out in Penning traps for precision measurement purposes, it is desirable to use linear Paul traps instead, leaning on the successes of trapped ion quantum processors.
Here we present a Paul trap for electrons made of a single printed circuit board. Our approach requires no assembly and the rigid design minimizes manufacturing intolerances.
We characterize the trap performance and observe trapped electron lifetimes of 2.13\,ms and secular frequencies of up to 90\,MHz.
\end{abstract}

\maketitle


\section{Introduction}
\label{sec:intro}

Trapped atomic ions are promising candidates for building a quantum processor with a large number of qubits \cite{Wineland1998,Blatt2008,Haeffner2008,Bruzewicz2019}.
Among their strengths are quantum state preparation, manipulation and readout with near unity efficiency, and long coherence and trap lifetimes, which allow for storing quantum information for a long time.
The implementation of quantum gates in trapped ions has been very successful, reaching single- and two-qubit fidelities of larger than $\,99.9$\% \cite{Harty2014,Ballance2016}. A remaining challenge comes with the large optical overhead such setups require when they are scaled up to many qubits.
Techniques that are explored include shuttling ions in QCCD architectures \cite{Kielpinski2002}, which employ separate zones for loading, storage and gate operations, and the use of laser-free gates \cite{Ospelkaus2008,Harty2016,Zarantonello2019}.

Qubits encoded in the spin of trapped electrons have been proposed and explored as an alternative to ion qubits \cite{Ciaramicoli2001,Ciaramicoli2003,Peng2017,Kotler2017,Daniilidis2013electron,Yu2022,Osada2022,Sutherland2022,Huang2025}. The use of electrons is expected to offer the many advantages of ions while removing the need for optical addressing. 
Moreover, there are various additional properties of electron qubits that make them a promising choice for future quantum platforms.
First, the two-level spin state of electrons is the simplest possible level structure for a qubit and it avoids unwanted leakage out of the computing Hilbert space \cite{Yu2022, Loss1998, Pla2013, Trifunovic2012}.
Second, since electrons are orders of magnitude lighter than atomic ions, the secular frequencies of trapped electrons can be significantly higher than those of trapped ions. This potentially enables faster qubit gates \cite{Sutherland2022, Huang2025, Ciaramicoli2003}.
Finally, coherent quantum operations on electrons are carried out using microwave and radio-frequency signals \cite{Li2026, Castoria2025, Chen2022, Li2024, Marzoli2009}, which simplifies the experimental architectures and replaces optical laser setups with robust, commercially available microwave technology.

Trapped electrons have been an established tool for studying quantum electrodynamics \cite{Sailer2022, Fan2023, Fan2025}, searches for physics beyond the standard model \cite{Carney2021}, and improved antihydrogen production
\cite{Walz1995,Mikhailovskii2026}.
While most experiments hitherto relied on Penning traps, using Penning traps for quantum information processing with electrons has encountered challenges and the confinement of a single electron was inhibited by anharmonicities in the trapping potential \cite{Bushev2008}.
As an alternate approach, individual electrons have been trapped on solid neon surfaces to be used as charge qubits \cite{Zhou2022,Zhou2023,Kanai2024,Zheng2025,Li2026,Li2026a}. The close proximity of a few nm to the surface, however, can enhance unwanted decoherence processes which have not been fully understood yet in such configurations \cite{Zhou2023}.
Following the first successful steering of electrons by a chip-based microwave guide \cite{Hoffrogge2011,Hammer2015,Zimmermann2021}, electrons have recently been confined in linear Paul traps \cite{Matthiesen2021,Taniguchi2025}.
Concurrently, efficient protocols for cooling and non-destructive detection and spin readout in such traps have been explored  \cite{Peng2017,Osada2022,Taniguchi2025}.

In this work, we present the confinement of electrons in a linear Paul trap made of a single printed circuit board.
The single body structure of this trap helps the system's mechanical stability and avoids alignment errors since the distance between the electrodes is fixed by the manufacturing process rather than assembled.
This aspect of our design can be particularly beneficial when traps are cooled down to cryogenic temperatures, which is necessary for implementing cooling and spin readout protocols.
Along the same lines, our approach offers better fabrication reproducibility since it avoids the need for complex assemblies.


\begin{figure*}[htpb]
    \centering
    \includegraphics[width=0.95\linewidth]{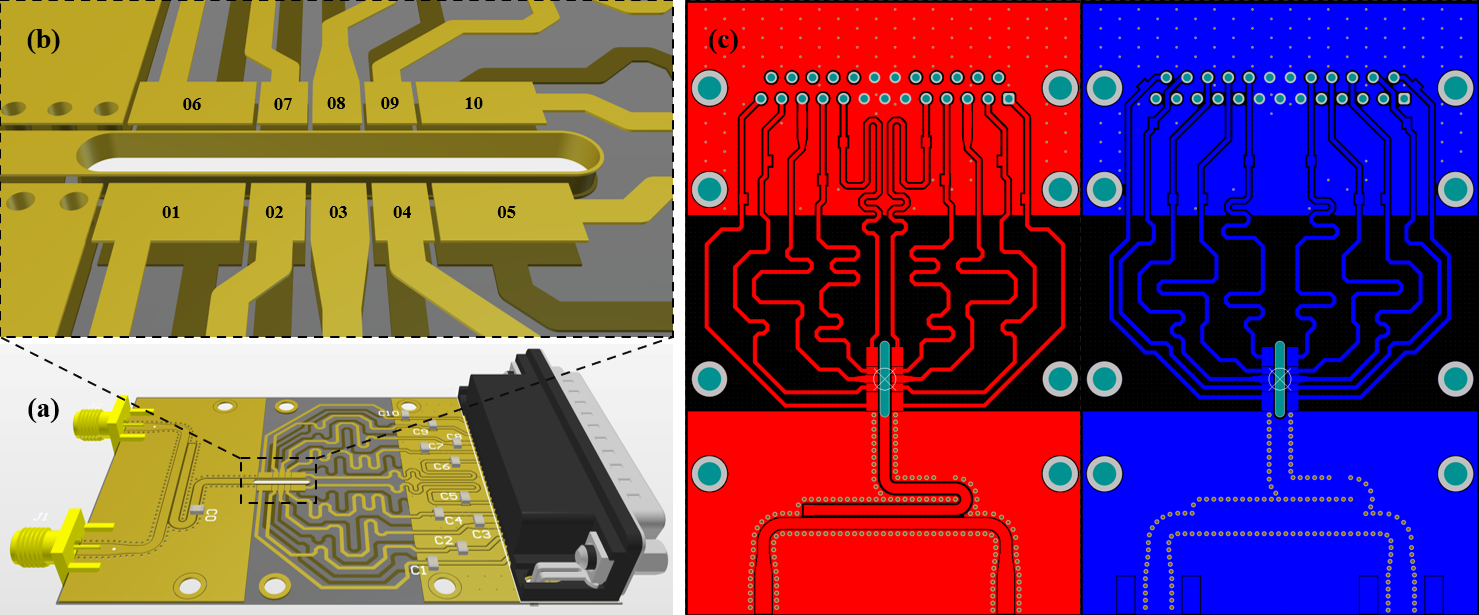}
    \caption{
    \textbf{(a)~3D model of the printed circuit board electron trap.} The RF signal for the Paul trap is delivered to the board by an SMA-connected feedline (left side of the board), which is capacitively coupled to an on-board RF resonator. Electrons are trapped near the end of the resonator, at the center of the elongated plated slit. The DC signals are delivered by a D-sub connector (right side of board).
    \textbf{(b)~Zoomed view of the slit.} The numbers 01 to 10 denote the DC electrodes on the top layer; a set of identical electrodes, 11 to 20, is located on the bottom layer. The slit is surrounded by a plated sidewall and a metal annular ring.
    \textbf{(c)~Schematics of the top (red) and bottom (blue) layers of the board.} The RF signal is delivered by a coplanar RF resonator capacitively coupled in parallel to a feedline on the bottom half of the board. The DC signal traces are routed such that all have an equal length. Each DC trace has added a capacitor that acts as a low-pass filter.}
    \label{fig:trap_layout}
\end{figure*}

This paper is organized as follows. In \rsec{sec:trap}, the trap geometry is presented, the PCB fabrication and the numerical potential simulation of the trap are discussed. \rsec{sec:exp_design} discusses the experimental setup and measurement procedure. In \rsec{sec:results}, a study on the secular trap frequencies and lifetime as a function of the trap parameters is presented.

\section{PCB Electron Trap}
\label{sec:trap}

Our electron trap is made of a single printed circuit board (PCB) with two conductive layers, as shown in \rfig{fig:trap_layout}. The PCB is fabricated out of a Rogers~RO4350B substrate with two 35~$\mu$m thick copper layers and finished with electroless nickel immersion gold (ENIG) plating ({\it Sierra Assembly}).

The board consists of two parts: a half-wave radio-frequency (RF) resonator and 10\,DC electrodes on each side.
Electrons are confined at the center of an elongated, side-plated slit cut into one end of the RF resonator.
For the following discussion, we define a right-handed coordinate system centered on the trap, with $\hat{x}$ normal to the board surface and pointing towards the top layer, $\hat{y}$ in the board plane along the width of the slot pointing from electrode 03 to electrode 08, and $\hat{z}$ along the slot pointing from electrode 05 to electrode 01.
The RF pseudopotential thus confines the electrons radially in the $\hat{x}$-$\hat{y}$ plane, while the combined DC fields provide axial confinement along $\hat{z}$.

\subsection{Trap Design}
\label{subsec:trap_design}

The RF subsystem consists of a half-wave coplanar resonator which is capacitively coupled in parallel to a feedline, both of which are designed as grounded coplanar waveguides (GCPW) with via fencing \cite{Ponchak2000}.
The feedline terminates in two SMA connectors soldered onto the trap board.
The RF drive enters from one of the SMA connectors, and part of its power is coupled to the resonator.
The residual power is transmitted and leaves the trap from the other SMA connector. This output is monitored by a spectrum analyzer (discussed in \rsec{subsec:control_system}).
The loaded quality factor of the resonator is $Q_L \approx 40.7$.

As shown in \rfig{fig:trap_layout}b, the slit that is cut into the resonator where the electrons are trapped is a slotted pad with a drill width of 48~mil bounded by an annular ring of 3~mil.
The width of the RF resonator electrode is 54~mil, which matches the width of the waveguide for a continuous transition to the trap region.
The slit sidewalls are plated, so the RF electrode wraps around the trap center on all sides and provides confinement in the two directions transverse to the slit.
A continuous ground plane on the back of the board (\rfig{fig:trap_layout}c) serves as the reference for the GCPW resonator and the feedline.
This ground plane breaks the symmetry between the two sides of the board normal to its surface, which results in a slight asymmetry of the radial pseudopotential (see \rsec{subsec:simulation}).

The DC subsystem consists of 20\,electrodes arranged in four rows, each containing five electrodes.
These electrodes are arranged symmetrically along the slit in both the top and bottom layers of the board, with one row on each side of the slit on each layer, as shown in \rfig{fig:trap_layout}b.
The ten electrodes lying on the top layer are labeled 01 to 10 and the ten on the bottom layer are labeled 11 to 20.
The DC electrodes are routed to a 25-pin D-sub breakout connector, and each trace has a ceramic capacitor of 10\,nF to filter high-frequency electrical noise.
The length of all DC traces is the same, so that the propagation delays across channels remain the same.
In addition to delivering voltages for axial confinement and stray-field compensation, these electrodes are also used to deliver an extraction pulse for electron detection and a radio-frequency signal for probing the secular motion of the electrons.
These signals are added to the DC voltages with bias tees.

The overall design of this trap leans on the assembled multi-PCB trap presented in \cite{Matthiesen2021}. Our approach minimizes potential geometrical asymmetries due to finite assembly tolerances. At the same time, an intrinsic drawback of our design is that the DC electrodes naturally need to be recessed from the RF electrode, and hence, from the trap center. Consequently, the effects of the DC fields are reduced by the loss of direct line of sight to the trap center and an increase in electron-electrode distance.
However, this drawback can easily be mitigated by amplifying the DC electric fields.

\subsection{Trap Potential Simulation}
\label{subsec:simulation}

To develop a theoretical model of our trap, we employed a finite-element solver HFSS \company{Ansys} to simulate the RF and DC electric fields for a given set of applied voltages. The simulation included a copper shield around the RF part of the trap (see \rsec{subsec:control_system}).

\begin{figure}[htbp]
    \centering
    \includegraphics[width=1\linewidth]{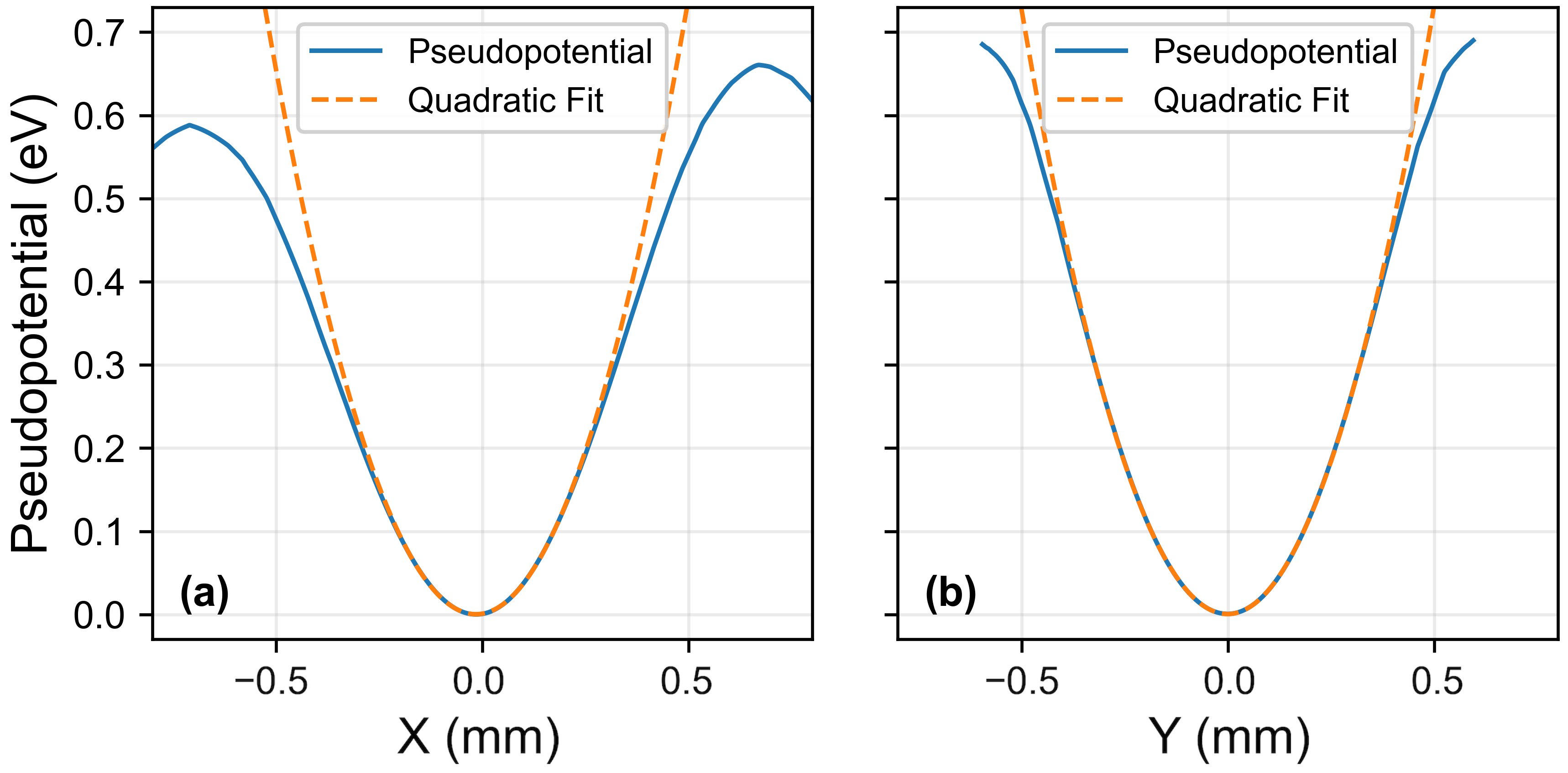}
    \caption{\textbf{Finite-element simulation of trap.}
    (a)~Pseudopotential along the $\hat{x}$-axis (perpendicular to the trap board) and (b)~along the $\hat{y}$-axis (along the slit width). 
    The fitted secular frequencies are $\omega_x = 2\pi \times 157.6$~MHz, $\omega_y = 2\pi \times 161.2$~MHz, and the trap depths are $U_x = 0.59$~eV and $U_y = 0.68$~eV.         
    All simulations assume an RF input power of 36\,dBm.
    }
    \label{fig:rf_simulation}
\end{figure}

The RF simulation predicts a resonance frequency of 1.753\,GHz for our resonator design.
The corresponding pseudopotential at this frequency with a RF drive of 36\,dBm applied at the feedline SMA connector is shown in \rfig{fig:rf_simulation}(a) and (b).
The simulated trap depth is 0.59\,eV along $\hat{x}$, and 0.68\,eV along $\hat{y}$ limited by the trap geometry.
A quadratic fit to the simulated pseudopotential near the trap center gives secular frequencies of $\omega_x = 2\pi \times 157.6$\,MHz and $\omega_y = 2\pi \times 161.2$\,MHz.
The pseudopotential is symmetric along $\hat{y}$ since the electrode layout is mirror-symmetric about the slit.
The slight asymmetry along $\hat{x}$ is a consequence of the single-sided ground plane for the GCPW.

The DC simulation is used to calculate the contribution of each individual DC electrode to the electric potential at the trap center. The result is then inverted to map the $\hat{z}$-direction potential in the trapping region onto the required voltage configurations of the DC electrodes \cite{Austin2010}.


\section{Experimental Setup}
\label{sec:exp_design}

\begin{figure}[htpb]
    \centering
    \includegraphics[width=1\linewidth]{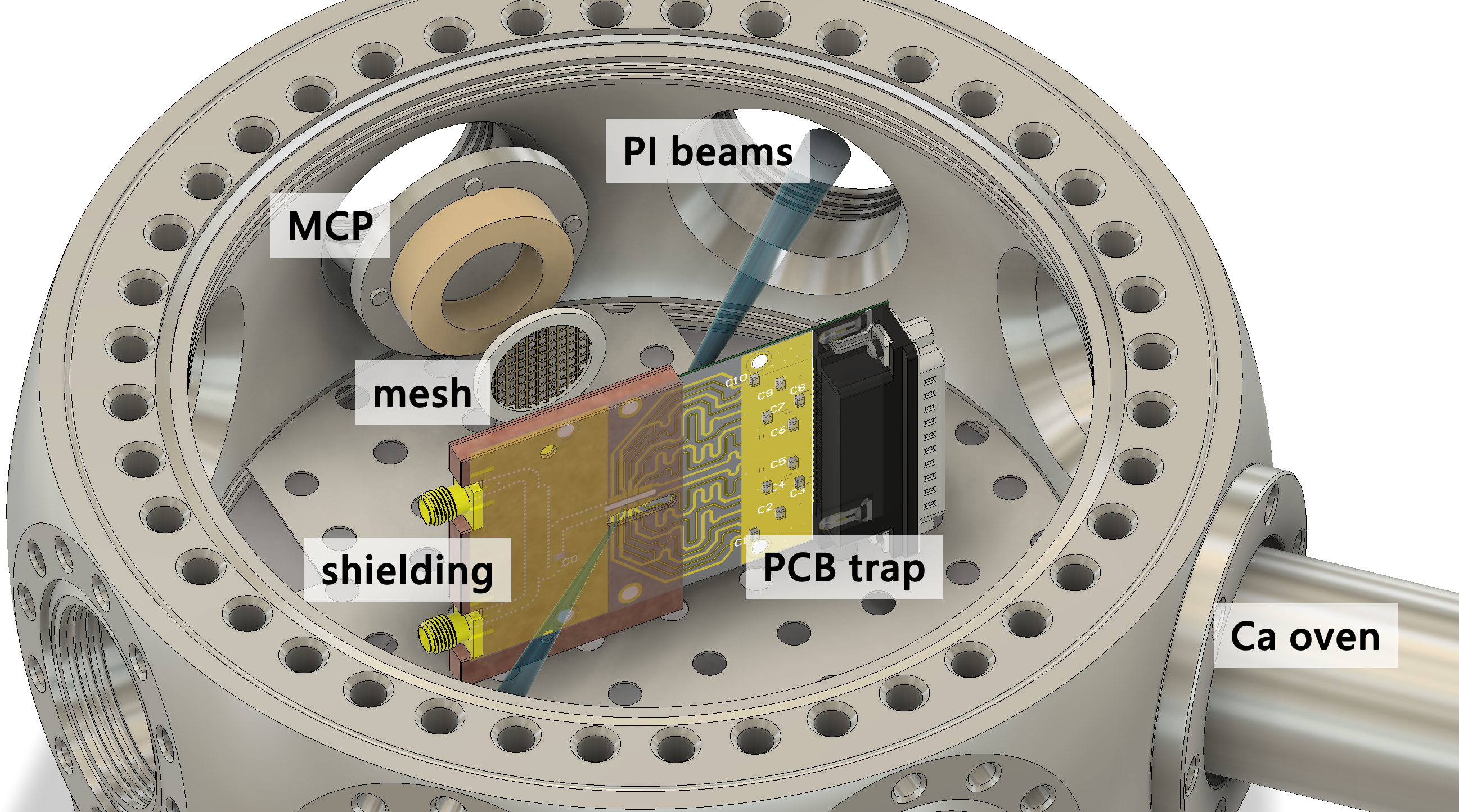}
    \caption{\textbf{Experimental apparatus}. The PCB Paul trap is mounted at the center of a UHV vacuum chamber, with the calcium oven (not visible), copper mesh, and microchannel plate detector arranged along the adjacent viewport axes.}
    \label{fig:vacuum_chamber}
\end{figure}

\rfig{fig:vacuum_chamber} shows a model of our experimental apparatus, which consists of a UHV vacuum chamber that is kept at a pressure of $\approx\,5 \times 10^{-8}$\,Torr using a turbo pump \company{Agilent, TwisTorr 305 FS} backed by a roughing pump \company{Agilent, IDP-7 scroll}.
The PCB trap is mounted in the center of the chamber.
The overall assembly is similar to the previous apparatus that used a multilayer trap \cite{Matthiesen2021}.
All experiments in this work are carried out at room temperature.

\subsection{Electron Source}
\label{subsec:electron_source}

To load electrons into the trap, the electrons must be created within the trapping region with a kinetic energy below the trap depth. From the various methods that have been proposed to create low-energy electrons \cite{Kotler2017}, we use photoionization of a thermal beam of atoms near the trap center since it has previously been successfully implemented to load a Paul trap for electrons \cite{Matthiesen2021}.

Specifically, a resistively-heated oven filled with calcium granules \cite{Gulde2001} generates an atomic beam which is oriented at an angle of $45^\circ$ to the trap axis and collimated by an aluminum iris attached to it. The calcium atoms are then ionized in a two-photon process by a 422\,nm laser \company{Toptica, DL Pro} and a 390\,nm laser \company{Moglabs, LDL}, which are aligned at an angle of $45^\circ$ with respect to the trap axis and perpendicular to the calcium beam and overlapped at the geometric trap center.
Specifically, the 422\,nm laser drives calcium atoms from their ground state $4s^2\,(^1$S$_0)$ to the first excited state $4s4p\,(^1$P$_1)$, and the 390\,nm laser further excites them to the continuum.
The combined energy of the two photons ($\approx\,6.11$\,eV) is close to the ionization threshold, and is tuned such that the free electron has a minimum kinetic energy in the meV regime, far below the trap depth.
The remaining calcium ion is not trapped since the trap parameters are optimized for the light electron mass and the trap is unstable for the heavy calcium ion.


\subsection{Trap Operation}
\label{subsec:control_system}

\begin{figure}[htpb]
    \centering    \includegraphics[width=1\linewidth]{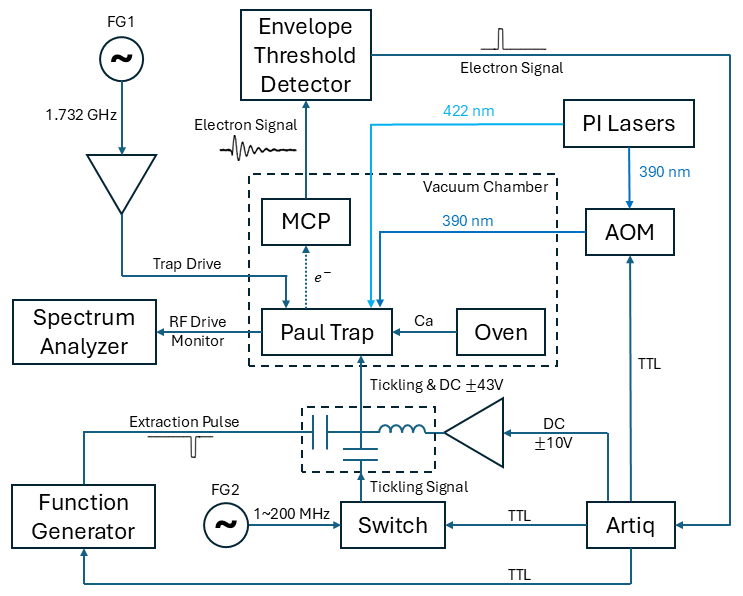}
    \caption{\textbf{Schematic diagram of the experiment}. The oven provides a thermal beam of calcium atoms, which are photoionized to produce trappable electrons. The Paul trap is driven at 1.732~GHz. Further details are explained in the main text.}
    \label{fig:system_flowchart}
\end{figure}

An overview of the control and detection system for the electron trap is shown in \rfig{fig:system_flowchart}.
The RF trap drive is generated by a signal generator \company{Rhode \& Schwarz, SMB-100A} and amplified through two cascaded stages \company{Mini-Circuits,  ZJL-5G+ and ZHL-5W-422+} to deliver a power of up to 36.8\,dBm at the trap's input SMA connector. The drive frequency is set to $\Omega_\textrm{rf} = 2\pi \times 1.732$\,GHz, matching the resonance of the on-board RF resonator.
The transmitted RF power is monitored by a spectrum analyzer \company{Keysight, N9000B} connected to the trap's output SMA connector.
Finally, to provide a better electrical ground, a copper box is added that shields the RF portion of the board.

The 20 DC electrodes on the trap board carry three types of signals: static voltages that provide axial confinement, negative voltage pulses that kick the electrons toward the MCP, and a weak RF excitation \textit{(tickle)} for probing the motional frequencies of the trapped electrons by resonantly increasing trap losses \cite{Vedel1990}.

The 20 channels of static voltages are generated by the DAC board of our experimental control system \artiq\ \cite{Artiq2021} in the range of $\pm 10$~V. 
Each channel is connected to a custom-build low-pass filter with a cutoff frequency of 33.8\,Hz to minimize RF noise on the DC lines. The DC signals are subsequently amplified by a custom-build non-inverting amplifier with a gain of 4.3, resulting in an expanded operating range of $\pm 43$~V.
On 6 of these 20 channels, 01, 03, 08, 10, 12, and 19, a bias-tee \company{Mini-Circuits, PBTC-1GW+} allows RF signals to be added to the DC signal. Specifically, electrodes 03 and 08 receive extraction pulses that eject electrons from the trap towards the MCP. Electrodes 01, 10, 12 and 19 are available for the tickling signal. The combined signal is delivered to the trap electrodes through a 25-pin D-sub ribbon cable.

For the photoionization process, the 422\,nm laser runs continuously during the full experiment cycle and is frequency stabilized to the reading of a wavelength meter \company{Bristol Instruments, 871} by a software PID. The 390\,nm laser is switched by an acousto-optic modulator (AOM) \company{Brimrose, TEF-85-20-390} and is operated free-running without frequency stabilization. We find experimentally that the photoionization efficiency is insensitive to frequency variations of the 390\,nm laser, likely due to the high density of Rydberg states in combination with AC Stark broadening and field ionization due to the RF field of the trap drive.

To detect electrons, a microchannel plate (MCP) \company{Hamamatsu, F2221S} is placed near the trap, facing the slit. The MCP is supplied with 400\,V at the front, 2400\,V at the back electrode, and 2600\,V at the anode \company{SRS, PS350}.
A custom-built copper mesh which is biased at $120$~V \company{Bertan Series 230} is placed in between the trap and the MCP to accelerate electrons towards the detector.
The electronic pulse the MCP generates upon the impact of an electron is first high-pass filtered to remove the DC bias of the anode voltage. The output is then low-pass filtered \company{Mini-Circuits, BLP-1200+} to suppress noise pickup from the RF trap drive and amplified \company{Mini-Circuits, ZFL-500LN} before being sent to an envelope threshold detector \company{Analog Devices, ADL5904}.
The last step converts the signal into a digital pulse that is sent to a TTL input of \artiq\ for edge counting.


\label{subsec:time_sequence}

The experimental cycle is divided into three phases: The first phase is the loading process during which the 390\,nm laser is switched on continuously to generate electrons and load them into the trap. 
The laser is then switched off for the (optional) waiting phase, during which no new electrons are produced. Finally, in the extraction phase, an arbitrary waveform generator \company{BK Precision, 4053B} applies a $-15$\,V pulse to electrodes 03 and 08, ejecting all remaining electrons from the trap and actively pushing them towards the MCP detector.
During each sequence, electrons escape the trap before the extraction pulse and a part of them end up at the MCP detector. To monitor these losses, we record three quantities with the MCP continuously during each sequence: the \textit{loading signal}, the electron counts during the loading phase; the \textit{lost signal}, the electron counts during the waiting phase; and the \textit{survival signal}, the electron counts during the extraction phase.


\section{Results}
\label{sec:results}

\subsection{Trapped Electron Lifetime}
\label{subsec:lifetime}

To characterize the trap performance, we measured the trapped electron lifetime.
Here, the loading process is set to 200\,\musec\ and the waiting times are varied from 0 to 5\,ms and the trapped electron counts are recorded.
For each point, a number of averages ranging from 500 to 2500 are taken and each measurement is repeated 100 times to build up statistics. The reason for increasing the number of averages is to address the weak survival signal for longer waiting times.

\begin{figure}[htbp]
    \centering
    \includegraphics[width=1\linewidth]{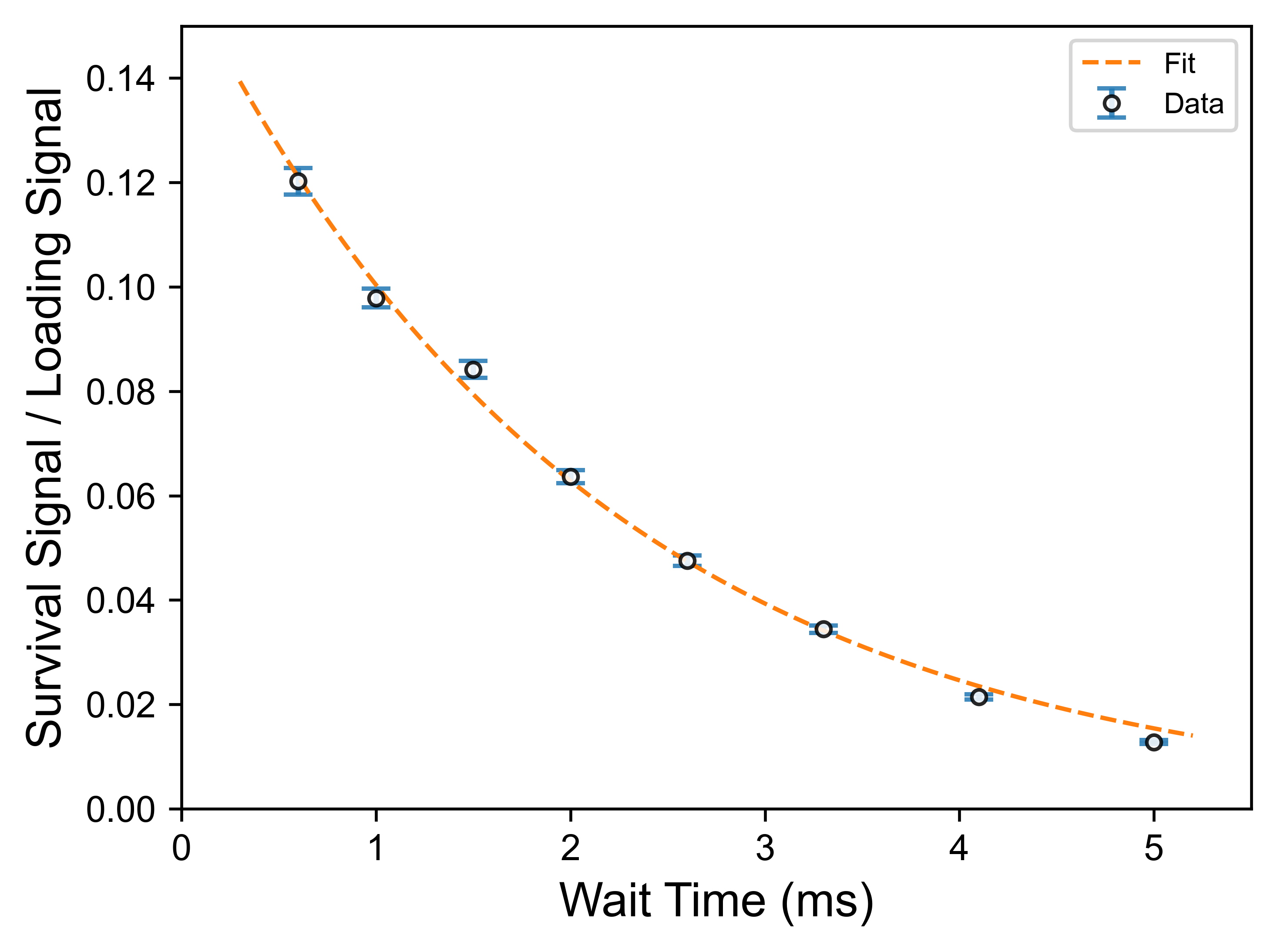}
    \caption{\textbf{Trapped electron lifetime measurement.}
    The data is fitted with an exponential decay, which yields a lifetime of 2.13\,ms.}
    \label{fig:lifetime}
\end{figure}

The measured survival signal normalized to the loading signal is shown in \rfig{fig:lifetime}.
We fit the data with a single exponential decay $\propto e^{-t/\tau}$, which yields a lifetime of $\tau$ = 2.13 $\pm$ 0.27\,ms.
This short lifetime stands in contrast to typically trapped ion lifetimes, which can be on the order of hours or even days. 

We attribute this limitation to various aspects: Imperfections in the electric fields due to finite manufacturing tolerances can lead to anharmonicities of the trapping potential. Furthermore, with the given setup, it is very challenging to fully compensate excess micromotion in the trap since the usual compensation techniques used in ion traps \cite{Berkeland1998} are not applicable.
Here, we systematically scanned the compensation fields while maximizing the trapped electron signal. However, this approach does not necessarily minimize excess micromotion.
Implementing more advanced techniques requires cooling and detection of the motion of electrons, for which a cryogenic system will be necessary \cite{Peng2017}.
Finally, we presume that the most likely reason for the short lifetime is due to the fact that the trapped electrons are not actively cooled in our room temperature setup.
This stands in contrast to trapped ions that are typically laser cooled to near the Doppler cooling limit. In our trap, the temperature of the electrons depends on where in the trap they are loaded and high temperature electrons describe large orbits and are more prone to exiting the trap when getting near the edge of the trapping potential.

\subsection{Motional Trap Frequencies}
\label{subsec:trap_freq}

\begin{figure}[htbp]
    \centering
    \includegraphics[width=1\linewidth]{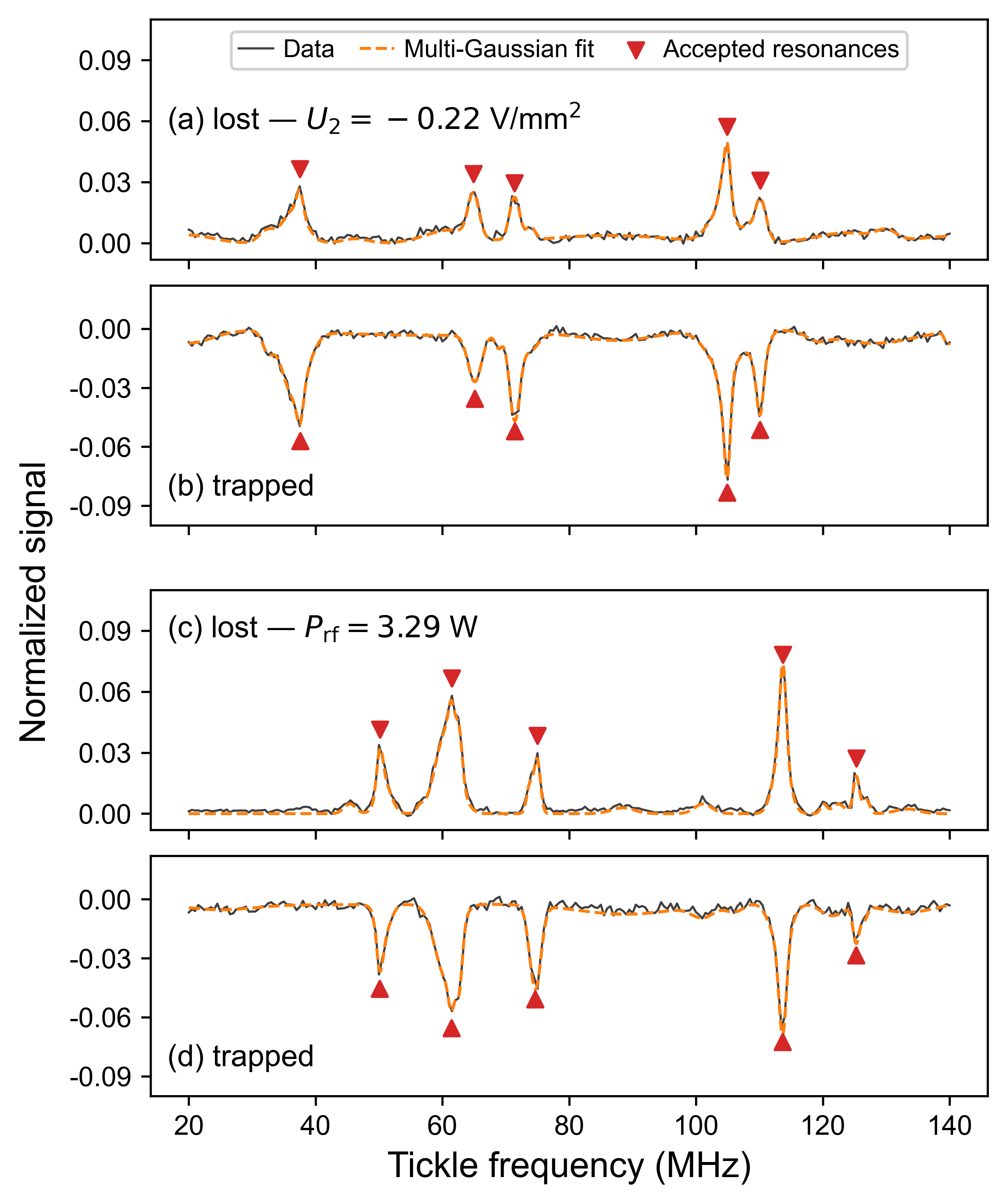}
    \caption{\textbf{Example tickling spectra and resonance extraction.}
    Normalized lost signal (lost signal / loading signal, panel a, c) and normalized trapped signal (surviving signal / loading signal, panel b, d) versus the tickle frequency, the data used here are the data after baseline correction.
	(a, b)~for a DC scan point at $U_2 = -0.22$~V/mm$^2$ (RF input power 3~W), and (c, d)~for an RF scan point at $P_\textrm{rf} = 3.29$~W ($U_2 = -0.35$~V/mm$^2$).
    The result shows that the dips of the trapped signal appear at the same frequencies as the peaks of the lost signal.}
    \label{fig:example_spectrum}
\end{figure}

\begin{figure*}[t]
    \centering
    \includegraphics[width=1\linewidth]{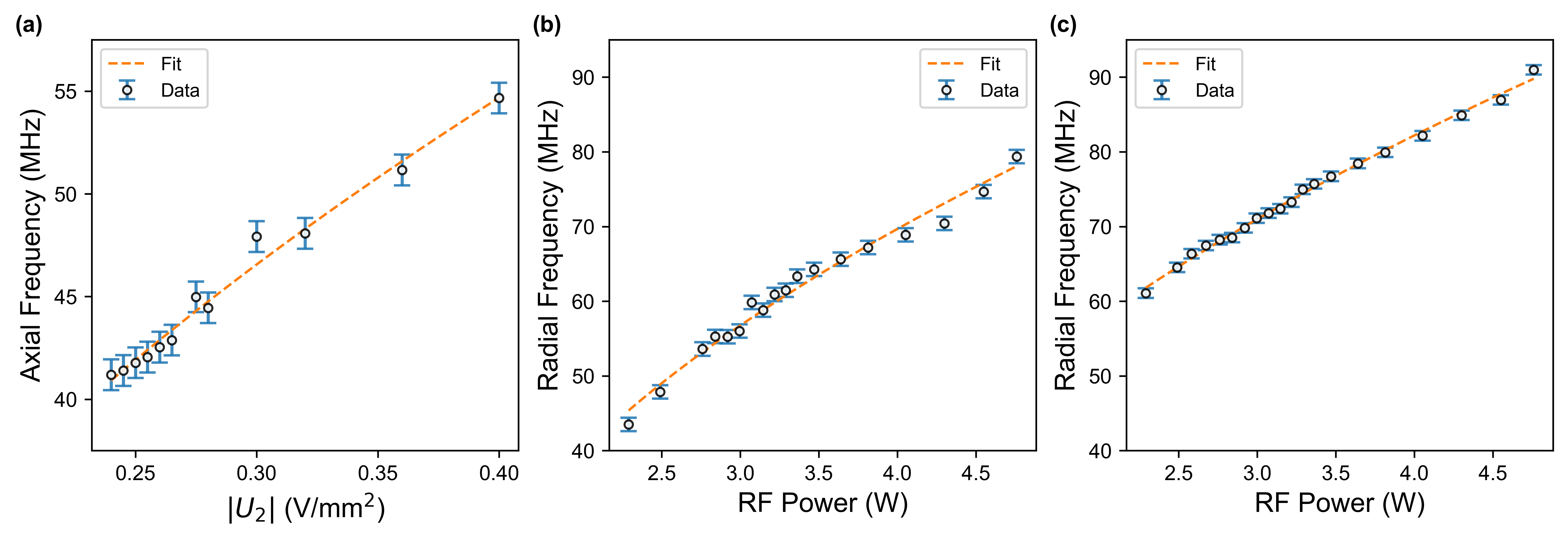}
    \caption{\textbf{Motional frequencies of the trapped electrons.}
    (a) shows the motional frequency of the trapped electrons in the axial direction as a function of $|U_2|$.
    (b) and (c) show the motional frequencies in the radial direction as a function of the RF input power.
    Dashed lines are fits to the data of \req{eq:wz_fit} and \req{eq:wxy_fit}.
    The error bars of the data represent the average linewidth of the measured resonance peaks.
    }
    \label{fig:motional_frequencies}
\end{figure*}

We further characterize the electron trap by measuring the secular axial and radial motional frequencies.
For this measurement, we use a loading time of 200\,\musec\ and a waiting time of 140\,\musec. During the waiting period of the sequence, the \textit{tickle} pulse is added to electrode 12 for a duration of 130\,\musec, activated 5\,\musec\ after the start of the waiting phase.

To probe the motional frequencies, we record the number of electrons that are lost as a function of the tickling signal frequency, ranging from 20 to 140\,MHz in steps of 0.5\,MHz steps.
Specifically, we measure the {\it survival} signal and the {\it lost} signal, both of which are normalized to the {\it loading} signal.
\rfig{fig:example_spectrum} shows two examples of the tickling spectra.
To each measured spectrum, an asymmetric least squares baseline correction \cite{Baek2015} with a smoothing parameter $\lambda = 200$ and an asymmetry parameter $p=0.01$ (for {\it lost} signal) / 0.99 (for {\it survival} signal) is applied.
Then an adaptive multi Gaussian peak fitting algorithm is applied to identify peaks on the spectrum.
This algorithm adds Gaussian peaks incrementally to the fitting function and exits when the coefficient of determination of the fit, $R^2\geq0.995$, or when the corrected Akaike information criterion \cite{Hurvich1989, Stoica2004} starts to increase.
From the fits, the peaks are selected that fulfill the following criteria: the relative amplitude is greater than 0.10, where the normalization is such that the minimum of the spectrum is 0 and the maximum of the spectrum is 1.0; the peak prominence is greater than 0.06; and the amplitude to width ratio is greater than 0.06.

This measurement is carried out as a function of the quadrupole amplitude of the DC potential ($U_2$) to probe the scaling of the axial motional frequencies and the RF input power ($P_\textrm{rf}$) to probe the scaling of the radial motional frequencies.
For each measurement, the resonances that consistently change position with a change of the RF or the DC potential are then identified as radial and axial resonances.
The experimental cycle is repeated 20,000 times for each point to build up statistics.
The measurement takes around one hour for each spectrum, and the entire experiment process over several days may be subject to environmental factors drifting over time. 
To mitigate this issue, for each $U_2$ or $P_\textrm{rf}$ value, we repeat the measurement 3 to 6 times in a random sequence and average the spectra accordingly. 
The results of the motional frequency measurements are shown in \rfig{fig:motional_frequencies}.
Part (a) shows the axial motional frequency as a function of $U_2$ for a fixed RF input power of 3\,W. We observe axial frequencies from $2\pi \times 40$ to $2\pi \times 55$~MHz as $|U_2|$ increases from 0.25 to 0.4\,V/mm$^2$.
Following \cite{Wineland1998}, we fit the result to
\begin{equation}
\label{eq:wz_fit}
    \omega_\textrm{axial}/2\pi = \alpha_\textrm{ax} \cdot \sqrt{|U_{2}| - o_\textrm{ax}}
\end{equation}
and find for $\alpha_\textrm{ax} = 90.68$\,MHz/($\sqrt{\textrm{V}}$/mm) and $o_\textrm{ax} = 0.04$\,V/mm$^2$.
Thus, we find a reasonable agreement with theoretical expectation of a square-root scaling with a negligible offset.

Part (b) and (c) show two different branches of the first order radial motional sideband. These measurements were taken at a fixed $U_2$ of $-0.35$\,V/mm$^2$. Here we observe frequencies ranging from $2\pi \times 44$ to $2\pi \times 78$\,MHz and $2\pi \times 60$ to $2\pi \times 90$\,MHz, respectively. At $\omega_\textrm{radial} = 2\pi \times 90$\,MHz, the corresponding stability parameter of the electron trap is $q = 2\sqrt{2} \omega_\textrm{radial}/\Omega_\textrm{rf} = 0.147$.
We fit the radial frequency data to
\begin{equation}
\label{eq:wxy_fit}
    \omega_\textrm{radial}/2\pi = \alpha_\textrm{ra} \cdot \sqrt{P_\textrm{rf} - o_\textrm{ra}}
\end{equation} 
which yields $\alpha_\textrm{ra} = 40.43$\,MHz/$\sqrt{\textrm{W}}$, $o_\textrm{ra} = 1.03$\,W for plot (b), and $\alpha_\textrm{ra} = 41.44$\,MHz/$\sqrt{\textrm{W}}$, $o_\textrm{ra} = 0.07$\,W for plot (c).


\label{app:secular_deviation}

\begin{figure}[htbp]
    \centering
    \includegraphics[width=1\linewidth]{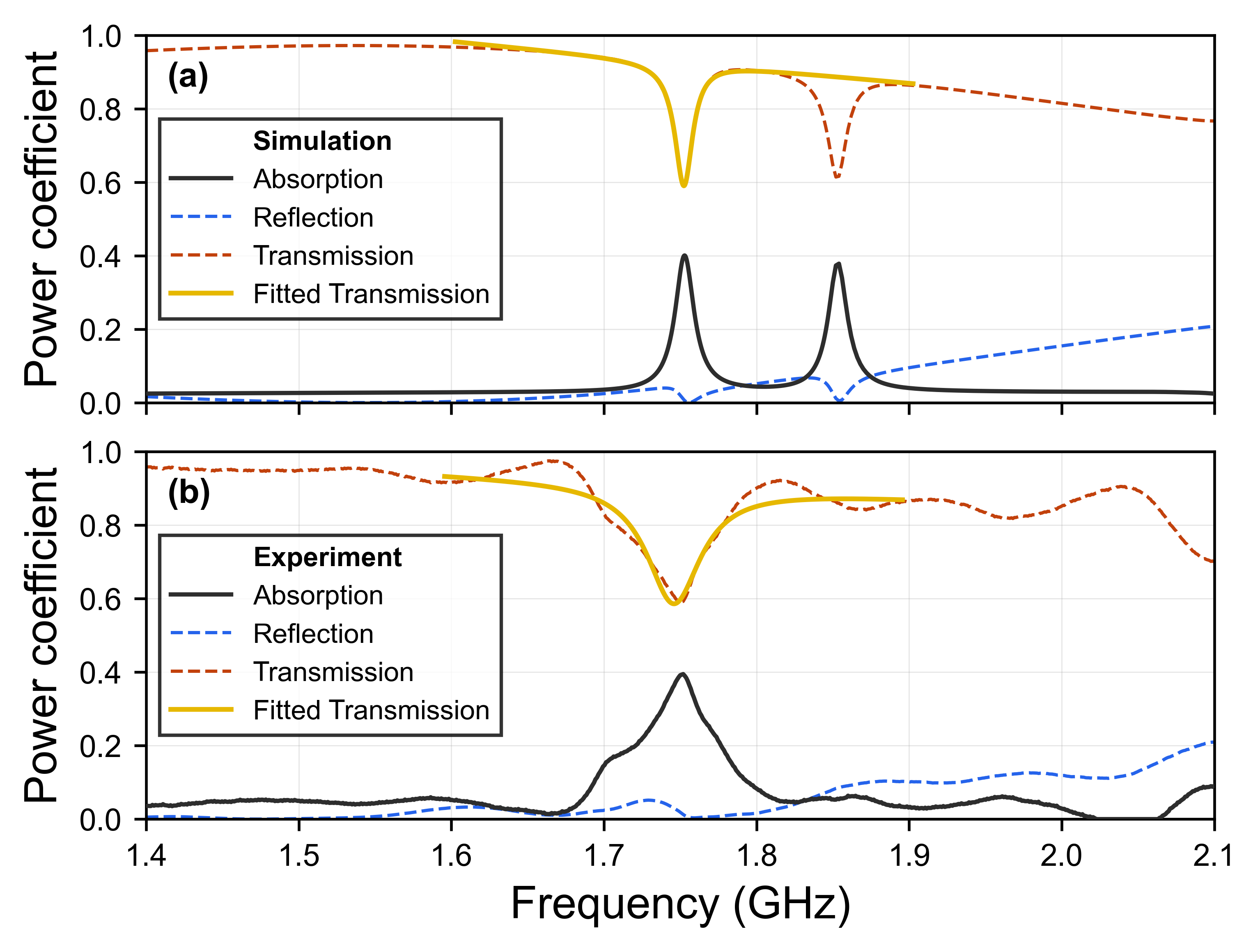}
    \caption{\textbf{Comparison of the (a)~simulated and (b)~experimentally measured reflection, transmission, and absorption coefficients of the trap.} the transmission (reflection) power coefficient $t=|S_{21}|^2$ ($r=|S_{11}|^2$) are extracted from the $S$-parameters measured between the two SMA connectors (shown in \rfig{fig:trap_layout}). The absorption coefficient is calculated using the conservation of energy $1-|S_{11}|^2-|S_{21}|^2$.}
    \label{fig:Q_factor}
\end{figure}

The measured and simulated radial trap frequencies (see \rfig{fig:rf_simulation}) differ by a factor of 2.2.
We attribute this deviation largely to non-optimal coupling to the RF resonator. To explore this limitation in more detail, we model the resonator as a hanger resonator coupled to a feedline \cite{McRae2020}. In this model, the  forward transmission coefficient $S_{21}$ of the RF transmission is described by
\begin{equation}
\label{eq:hanger_model}
    S_{21}(\omega) = g(\omega) \cdot e^{\ic(\alpha - \omega \tau)}
    \left(1 - \frac{Q_L\, e^{\ic\varphi}/Q_c}{1 + 2 \ic Q_L\!\left(\omega/\omega_0 - 1\right)}\right)
\,,
\end{equation}
where $Q_{c,i,L}$ is the coupling, internal, and loaded quality factor of the resonator, following $1/Q_L=1/Q_i+1/Q_c$, $\omega_0$ is the resonance frequency, $\tau$ is the transmission line delay coefficient and $\alpha$ is the initial phase, and $\varphi$ is the asymmetry factor.
To account for the frequency dependent transmission line loss, we introduce a linear magnitude correction $g(\omega) = a - b\,\omega/2\pi$ to the model. By fitting \req{eq:hanger_model} to the complex-valued simulated and measured $S_{21}$ (shown as power coefficients in \rfig{fig:Q_factor}), we extract the resonance frequency $\omega_0$, quality factors $Q_L, Q_c, Q_i$, and asymmetry parameter $\varphi$, listed in \rtab{tab:Q_fit}.

To relate the fitted $Q$ factor to the secular frequency, we note that for an on-resonance hanger resonator, the stored energy as a function of the input power $P_\textrm{rf}$ and the resonance frequency $\omega_0$ is given by \cite{McRae2020, Khalil2012}
\begin{equation}
    E_{\textrm{res}} 
    = 
    2 Q_L ^2 
    \dfrac
    {P_\textrm{rf}}
    {\omega_0 Q_c}
    \label{eq:resonator_energy}
\end{equation}

Under the pseudopotential approximation, the trapping motional frequency is proportional to the rf voltage amplitude at the resonator, $\omega_r \propto U_\textrm{rf} \propto \sqrt{E_\textrm{res}}$. Combining this with \req{eq:resonator_energy} yields $\omega_r\propto Q_L/\sqrt{Q_c}$. Therefore,
\begin{equation}
    \dfrac
    {\omega_{r,\textrm{exp}}}{\omega_{r,\textrm{sim}}}
    =
    \dfrac
    {Q_{L,\textrm{exp}}}
    {Q_{L,\textrm{sim}}}
    \sqrt{
    \dfrac
    {Q_{c,\textrm{sim}}}
    {Q_{c,\textrm{exp}}}
    }
    =    
    0.565
    \,.
\end{equation}
Applying this correction to the simulation values (see \rfig{fig:rf_simulation}), the adjusted radial frequencies are $\tilde{\omega}_{x} \approx 2\pi \times 89.0$\,MHz and $\tilde{\omega}_{y} \approx 2\pi \times 91.1$\,MHz, which are significantly closer to the measured values.

\begin{table}[h]
    \centering
    \caption{Fitted Resonator Parameters}
    \begin{tabular}{lrr}
        \hline
        Parameter & Simulation & Experiment \\
        \hline
        $\omega_0/2\pi$ (GHz) & $1.753$    & $1.746$\\
        $Q_L$        & $124 \pm 1$         & $40.7 \pm 0.9$       \\
        $Q_c$        & $618 \pm 2$         & $207 \pm 4$          \\
        $Q_i$        & $156 \pm 1$         & $51 \pm 2$           \\
        $a$          & $1.31 \pm 0.01$   & $1.14 \pm 0.01$    \\
        $b$          & $0.020 \pm 0.003$   & $0.109 \pm 0.004$    \\
        $\alpha$ (deg)  & $16.1 \pm 0.4$   & $4.8 \pm 0.4$        \\
        $\tau$ (ns)  & $0.307 \pm 0.001$   & $0.336 \pm 0.001$    \\
        $\varphi$ (deg) & $-2.4 \pm 0.2$   & $1.0 \pm 1.0$        \\
        \hline
    \end{tabular}
    \label{tab:Q_fit}
\end{table}

Finally, we note that the observed frequency difference between the two radial modes is a factor of 3-4 larger than what our simulation predicts. We suspect that this deviation is caused by asymmetries in the trap geometry and corresponding trap anharmonicities, which lift the radial-mode degeneracy further.


\section{Conclusion}
\label{sec:conclusion}

In this work, we have presented a linear Paul trap for confining electrons made of a single printed circuit board with two conductive layers.
We have demonstrated successful electron trapping with lifetimes of 2.13\,ms and observed secular trapping frequencies of up to $2\pi \times 55$\,MHz (axially) and $2\pi \times 90$\,MHz (radially).
Since the trap is manufactured as a single component it avoids imprecision of an assembly and supports reproducibility when attempting to manufacture multiple equal traps. For the same reason, we expect that our approach offers an improved cryogenic compatibility, which we plan to test as a next step.
Beyond quantum computing, this novel structure can also benefit positron and positronium research, electron optics, and plasma studies.


\begin{acknowledgments}
Z.~L., J.~.E.~, T.~W.~and B.~H.~acknowledge funding from the AFOSR through Grant No.~FA9550-24-1-0337. This material is based upon work supported by the Air Force Office of Scientific Research under award number FA9550-21-1-0427. This research was supported in part by grant NSF PHY-2309135 to the Kavli Institute for Theoretical Physics (KITP).
We would like to thank Hartmut Haeffner and the Haeffner group at UC Berkeley for useful discussions.
\end{acknowledgments}


\bibliography{library_processed}

\end{document}